\documentclass[prb,twocolumn,floatfix,superscriptaddress,showpacs]{revtex4}

\usepackage{graphicx}
\usepackage{epsfig}
\usepackage{dcolumn}
\usepackage{amssymb}
\usepackage{bm}

\begin{document}



\title{Negativity of the excess noise of a quantum wire capacitively coupled to a  gate}

\author{F. Dolcini}
\affiliation{Scuola Normale Superiore \& NEST-INFM-CNR, 56126
Pisa, Italy}
\author{B. Trauzettel}
\affiliation{Instituut-Lorentz, Universiteit Leiden, 2300 RA
Leiden, The Netherlands} \affiliation{Department of Physics and
Astronomy, University of Basel, 4056 Basel, Switzerland}
\author{I. Safi}
\affiliation{Laboratoire de Physique des Solides, Universit\'e
Paris-Sud, 91405 Orsay, France}
\author{H. Grabert}
\affiliation{Physikalisches Institut,
Albert-Ludwigs-Universit\"at, 79104 Freiburg, Germany}

\date{October 2006}

\begin{abstract}
The electrical current noise of a quantum wire is expected to
increase with increasing applied voltage. We show that this
intuition can be wrong. Specifically, we consider a single channel
quantum wire with impurities and with a capacitive coupling to
nearby metallic gates and find that its excess noise, defined as
the change in the noise caused by the finite voltage, can be
negative at zero temperature. This feature is present both for
large ($c \gg c_q$) and small ($c \ll c_q$) capacitive coupling,
where $c$ is the geometrical and $c_q$ the quantum capacitance of
the wire. In particular, for $c \gg c_q$, negativity of the excess
noise can occur at finite frequency when the transmission
coefficients are energy dependent, i.e. in the presence of
Fabry-P\'erot resonances or band curvature. In the opposite regime
$c \lesssim c_q$, a non trivial voltage dependence of the noise
arises even for energy independent transmission coefficients: at
zero frequency the noise decreases with voltage as a power law
when $c < c_q/3$, while, at finite frequency, regions of negative
excess noise are present due to Andreev-type resonances.
\end{abstract}

\maketitle

\section{Introduction}

In equilibrium, the fluctuations of an observable are directly
connected to the relaxation of its average value through the
fluctuation-dissipation theorem \cite{CW}. In contrast, out of
equilibrium fluctuations contain in general more information than
the mere average values. Electron transport is a typical example:
When a conductor is driven out of equilibrium by applying a finite
bias voltage $V$, the frequency spectrum\cite{Leso-Loos-JETP97}
\begin{eqnarray} \label{noise}
S(x,\omega,V) = \int_{-\infty}^{\infty} dt e^{i\omega t}
\left\langle \Delta j (x,t) \Delta j(x,0) \right\rangle \;
\end{eqnarray}
of the electrical current fluctuations $\Delta j(x,t) = j(x,t) -
\langle j(x,t) \rangle$ cannot be directly related to the ac
conductivity $\sigma(x,\omega)$. Here $\langle \ldots \rangle$
denotes the quantum-statistical average over the stationary
non-equilibrium density matrix of the system in presence of the
applied voltage $V$.

The expression (\ref{noise}) is the unsymmetrized version of the
customarily defined \cite{Landau,Kogan} current noise, the latter
being easily obtained as $S_{\rm
sym}(x,\omega,V)=(1/2) [ S(x,\omega,V)+S(x,-\omega,V) ]$. The
unsymmetrized noise has recently attracted attention in mesoscopic
physics, in that it can be directly measured by on-chip detectors,
as proposed in Refs.~\onlinecite{Leso-Loos-JETP97} and
\onlinecite{Aguado}, and experimentally realized for the first
time in Ref.~\onlinecite{deblo03}. In this experiment the noise is
related to the photon-assisted current generated in a
superconductor-insulator-superconductor (SIS) Josephson junction
detector.

Differently from the DC current, the noise is in principle a
position-dependent quantity\cite{blanter00}; the coordinate $x$ in
Eq.~(\ref{noise}) can be associated with the point of measurement
typically in the leads. The estimate of the actual value of the
distance~$\delta_x$ between $x$ and the lead-wire contact depends
on the device, the geometry of contacts, and the type of detector,
and therefore goes beyond the purposes of the present work. Here
we rather emphasize that, while at zero frequency the noise
becomes independent of the position, at finite frequency a
dependence on $x$ arises\cite{Trauz-Gra}. Since in this paper
finite frequency noise is discussed, we keep the dependence on $x$
explicitly.

The non-equilibrium noise provides insight to the mechanisms
underlying electron transport. In the shot noise limit  the
differences with respect to the equilibrium situation are most
significant. In particular, at zero frequency and zero temperature
the equilibrium noise vanishes, whereas the non-equilibrium noise
is in general finite. For instance, in a ballistic conductor with
impurities the shot noise originates from the statistical
transmission or reflection of the discrete charge carriers at the
scatterers. If the electron-electron interaction can be neglected,
the scattering matrix formalism \cite{blanter00} is applicable. In
this case the shot noise at zero temperature
reads\cite{Khlus87,Leso89,MBPRL90}
\begin{equation}
S(\omega=0,V) = \frac{e^2}{2 \pi \hbar} e V \, \sum_n T_n (1-T_n) ,
\label{shot-LB}
\end{equation}
where $T_n$ are the transmission coefficients of the eigenchannels
of the conductor, and $e>0$ is the elementary charge. Note that,
as mentioned above, the shot noise is independent of the position
$x$ appearing in Eq.~(\ref{noise}). Comparing Eq.~(\ref{shot-LB})
with the expression
\begin{equation}
G=\frac{e^2}{2 \pi \hbar} \sum_n T_n \quad \label{G-LB}
\end{equation}
for the conductance, one can see explicitly that, in view of the
$(1-T_n)$ suppression factors, the out of equilibrium noise cannot
be expressed in terms $G$.

In comparison with the formidable efforts made lately to predict
\cite{engel04,trauz04,lebed04,Pham,Pisto} and measure
\cite{schoe97,deblo03,Onac1,deblo06,Onac2} the frequency
dependence of the noise, not so much interest has been devoted to
the investigation of the {\it voltage} dependence of the noise. Intuitively, one would expect that the noise
should increase with the voltage $V$; this intuition is confirmed
by Eq.~(\ref{shot-LB}) describing the simple case of a system of
non-interacting electrons when both the energy dependence of the
transmission coefficients and the band curvature can be neglected.

This intuition is, however, wrong in general. For a single-channel
wire, for instance, Lesovik and Loosen\cite{Leso-Loos} have shown
that in the particular case where the transmission coefficient is
non-vanishing only in an energy window $\delta E$, the Fermi
energy lies within this energy window, and the temperature is
sufficiently high ($k_B T \gg \delta E$), the shot noise in the
regime $eV \gg k_B T$ is smaller than the equilibrium noise. In
the opposite limit of a multichannel clean wire it has been
demonstrated\cite{Bula-Rubi} that, although the {\it current}
noise is always increasing with $V$, the {\it voltage} noise may
decrease with bias. These results, however, are concerned with
either the high temperature regime or the semiclassical limit
(number of channels tending to infinity) of transport. Since most
of the experimental interest in mesoscopic conductors lies instead
in quantum effects, the open question is whether similar behavior
can occur in the deep quantum regime, i.e. for a finite number of
channels at low temperatures, where $\hbar \omega \gg k_B T$. This
paper aims at investigating this problem. We consider here a
single channel quantum wire at zero temperature. For simplicity,
we discuss the spinless case, although our analysis can be easily
generalized to spinful electrons. We analyze the conditions under
which the noise, both at zero and at finite frequency, can
decrease with bias, and investigate, in particular, whether the
excess noise
\begin{eqnarray}
S_{\rm EX}(x,\omega,V)&=&S(x,\omega,V)-S(x,\omega,0) , \label{SEX}
\end{eqnarray}
characterizing the change of the noise due to the finite voltage
with respect to the equilibrium case, can be negative. The total noise $S(x,\omega,V)$ is, of course,  always positive, as follows from the Wiener-Khintchine theorem\cite{WK}. A negative excess noise simply means that driving the system out of equilibrium by applying a voltage~$V$ reduces the noise in certain frequency regions with respect to its equilibrium value.\\

The noise spectrum (\ref{noise}) can be directly related to the
current spectral density
\begin{equation}
j(x,\omega)=\int_{-\infty}^{\infty}  dt\, e^{i \omega t} \, j(x,t)
\end{equation}
through the relation
\begin{equation}
S(x,\omega,V)=\frac{1}{2 \pi} \int_{-\infty}^{+\infty}
d\omega^{\prime} \langle \Delta j(x,\omega) \Delta
j(x,\omega^{\prime}) \rangle  \label{noise-jomega}
\end{equation}
where $\Delta j(x,\omega)=j(x,\omega)-\langle j(x;\omega)
\rangle$. As a consequence of the continuity equation the net flux
of current flowing into the conductor equals the time rate of
change of the charge in the conductor. Explicitly,
\begin{equation}
j(\frac{L}{2};\omega)-j(-\frac{L}{2};\omega) =
\int_{-\frac{L}{2}}^{\frac{L}{2}} dx  \, \mbox{div} j(x;\omega) =
i \omega Q(\omega) \label{cont-eq}
\end{equation}
where $x=\pm L/2$ are the locations of the edges of the conductor
of length $L$, and
\begin{equation}
Q(\omega)=\int_{-\infty}^{\infty}  dt\, e^{i \omega t}
\int_{-\frac{L}{2}}^{\frac{L}{2}}  dx  \, \rho(x,t)
\end{equation}
with the charge density operator $\rho(x)$ .

The wire is capacitively coupled to its electromagnetic
environment consisting of other metallic conductors nearby. This
gives rise to a {\it geometrical} capacitance~$C$ of the wire to
the environment. The right hand side of Eq.~(\ref{cont-eq}) can
then be interpreted as the displacement current
\begin{equation}
j_d(\omega)= i \omega Q(\omega)\quad   \label{iomegaQ-jd}
\end{equation}
through this capacitance~$C$. Furthermore, the capacitive coupling
induces a fluctuating shift $\Delta U(\omega) \sim Q(\omega)/C$ of
the band-bottom of the wire which modifies the energy of the
electrons, affecting in turn their scattering processes inside
the conductor and therefore the current $j(x,\omega)$ itself. This
feed-back process makes the problem of determining finite
frequency transport properties an
essentially {\it interacting} problem\cite{Blanter,SafiEPJB,blanter00}. In a DC-biased set-up,  the average current $I=\langle j(x,t) \rangle$ is independent of the position $x$ and the time $t$, so that the {\it average} current spectral density is simply $\langle j(x,\omega)\rangle =  2 \pi  I \delta(\omega)$, and the displacement current has vanishing expectation value, as can be seen from Eqs.~(\ref{cont-eq}) and (\ref{iomegaQ-jd}). For the noise, however, the full frequency-dependence of the {\it fluctuations} of the current spectral density plays a role (see Eq.(\ref{noise-jomega})), and the band-bottom shift induced by the displacement current fluctuations has to be taken into account. \\
\begin{figure}
\vspace{0.3cm}
\begin{center}
\epsfig{file=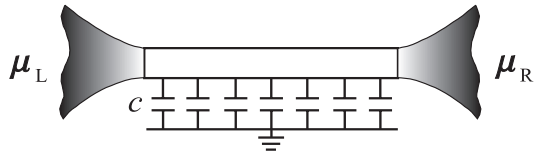,height=2.0cm,width=7.0cm}
\caption{\label{Fig1} Schematic of a quantum wire capacitively
coupled to a nearby gate (with a capacitance per unit length $c$).
The wire of finite length $L$ is connected at $x=\pm L/2$ to
reservoirs with electrochemical potentials $\mu_L$ and $\mu_R$. }
\end{center}
\end{figure}

In the present paper, we shall analyze the finite frequency noise
for the case of a ballistic single-channel quantum wire,
capacitively coupled to a gate (see Fig.~\ref{Fig1}). For
simplicity, we shall treat the capacitance $C$ as uniformly
distributed along the wire, $C=c L$, where $c$ is the capacitance
per unit length, and $L$ the length of the wire\cite{Foot1}. The
Hamiltonian of the system thus reads
\begin{equation}
H=H_{0} + H_{\rm imp} + H_{c} \label{Htot}\, .
\end{equation}
Here, $H_{0}$ describes the band kinetic energy of the electrons
in the wire, $H_{\rm imp}$ the scattering with impurities if
present, and \begin{equation} H_{c}= \frac{1}{2c}
\int_{-L/2}^{L/2} \delta\rho^2(x) \, dx \label{HC}
\end{equation}
the capacitive coupling to the gate, where
$\delta\rho(x)=\rho(x)-\rho_F$ is the deviation of the electron
charge density $\rho(x)$ from the density $\rho_F$ of an
electroneutral wire. The capacitance $c$ is a crucial quantity to
determine the finite frequency noise, for it relates the
displacement currents to the band-bottom shift $\Delta U$ in the
wire. In the regime   $c \gg c_q$, where $c_q=e^2 \nu$ is the quantum
capacitance and~$\nu$~is the density of states
per unit length of the electron band described by~$H_0$, it is
evident that a finite displacement current yields a very small
shift~$\Delta U$. The term~(\ref{HC}) is then negligibly small
with respect to~$H_0$, and therefore the problem of determining
the particle current decouples from the evaluation of the
displacement current. For a finite capacitance, however, the two
problems have to be solved simultaneously, and the band-bottom
shift cannot be neglected. In this paper we investigate this
problem. We are interested in the effects of the capacitive
coupling between wire and metallic gate, and not in the effects
arising from variations of the gate potential~$V_g$; we therefore
assume for definiteness that the gate is grounded~$V_g=0$.

The paper is organized as follows: In Sec.~\ref{Sec2}, we analyze
the case of a large geometrical capacitance (which corresponds to
the non-interacting problem). In this case, we show that in a
quantum wire with poorly transmitting contacts to the leads the excess noise can
be negative at finite frequency. In Sec.~\ref{Sec_Lut}, we analyze
the case of arbitrary geometrical capacitance. Thus, we discuss
the fully interacting problem. In particular, we adopt a mapping
between a capacitively coupled quantum wire and a Luttinger liquid
model to predict that the excess noise can be negative even in
presence of just a single impurity in the wire. Finally, we
conclude in Sec.~\ref{Sec_con}.

\section{The case of large capacitance}
\label{Sec2}

We first consider the case of large capacitance between wire and
gate ($c \gg c_q$). Then the displacement current (which is always
present to ensure charge conservation) does not induce a
band-bottom shift in the wire. The term (\ref{HC}) can then be
neglected and displacement currents play no role. We shall see
that some interesting effects can still arise from band curvature
or an energy dependence of the transmission coefficients which
shall therefore be retained. For a Hamiltonian
\begin{equation}
H_0=-\frac{\hbar^2}{2 m} \int_{-\infty}^{+\infty} \Psi^{\dagger}(x)
\partial^2_x \Psi(x) \, dx \label{H0par}
\end{equation}
with the usual parabolic dispersion, the expression for the
electronic current reads
\begin{equation}
j(x) = \frac{e \hbar}{2m i} \left\{\Psi^{\dagger}(x)
\partial_x \Psi^{}(x)- (\partial_x \Psi^{\dagger}(x) )\, \Psi^{}(x) \right\} , \label{j-PAR}
\end{equation}
where $\Psi(x)$ is the electron field operator. However, when the
band dispersion is linearized to investigate the low energy limit,
or when a discrete lattice model is considered, the expression
(\ref{j-PAR}) has to be consistently modified in order to preserve
the continuity equation for the solutions. For a Hamiltonian with
linearized spectrum \begin{eqnarray} H_0 = &-& i \hbar v_F \int
\Bigl( : \Psi^\dagger_{\rightarrow}(x)
\partial_x \Psi_{\rightarrow}(x) : \nonumber  \\
& & \hspace{1cm}  -: \Psi^\dagger_{\leftarrow}(x) \partial_x
\Psi^{}_{\leftarrow}(x) : \Bigr) dx , \label{H0}
\end{eqnarray}
the expression for the current reads
\begin{equation}
j(x)=e v_F \left(: \Psi^{\dagger}_{\rightarrow}(x)
\Psi^{}_{\rightarrow}(x) - \Psi^{\dagger}_{\leftarrow}(x)
\Psi^{}_{\leftarrow}(x) : \right)  . \label{j-DIR}
\end{equation}
Here, $\Psi_{\rightarrow}$ ($\Psi_{\leftarrow}$) is the electron
field operator for a right (left) moving particle, and the symbol
$: \, \, :$ denotes normal ordering with respect to the Dirac
sea\cite{vonDelft}. Similarly, in a tight-binding lattice model,
\begin{equation}
H_0=-\tau \sum_{i} \left( c^{\dagger}_i c^{}_{i+1} +
c^{\dagger}_{i+1}
  c^{}_{i} \right) , \label{H0TBM}
\end{equation}
one obtains
\begin{equation}
j(x)=-\frac{e \tau}{\hbar} \left( c^{\dagger}_i c^{}_{i+1} -
  c^{\dagger}_{i+1} c^{}_{i} \right)  , \label{j-TBM}
\end{equation}
where $\tau$ is the hopping energy, $c_i$ annihilates a fermion on
lattice site $i$, and $x=i a_0$ (with $a_0$ the lattice spacing
and $i$ the lattice index).

All these different types of models can be treated within the
scattering matrix formalism\cite{blanter00}. Here we follow the
standard notation and denote by $a_{X E}$ ($b_{X E}$) the
operators of incoming (outgoing) states of energy $E$ at the side
$X=R,L$ of the scatterers; the outgoing states are expressed in
terms of the incoming states through the $\mathsf{S}$-matrix:
\begin{equation}
\left(
\begin{array}{l}
 b_{L E} \\
 b_{R E}
\end{array} \right) = \mathsf{S}(E) \left(
\begin{array}{l}
a_{L E} \\
a_{R E}
\end{array} \right) ; \hspace{0.3cm} \mathsf{S}(E)=\left(
\begin{array}{cc}
 r^{}(E) & t^\prime(E) \\
 t^{}(E) & r^\prime(E)
\end{array} \right) . \label{Smatr}
\end{equation}
The electron operator $\Psi$ in the left lead can be written as a
superposition of $a_{L E}$ and $b_{L E}$ for all energies,
weighted by the related eigenfunctions $\alpha_{L E}(x)$ and
$\beta_{L E}(x)$ of incoming and outgoing states (typically plane
waves). The cases of a linearized spectrum and of a lattice model
can be treated in the same way. Substituting the above energy mode
expansion into the expressions
 (\ref{j-PAR}), (\ref{j-DIR}) or (\ref{j-TBM}), and using Eq.~(\ref{Smatr}),  the time evolution of the current operator at a point $x$ located in the  left lead reads \cite{blanter00}
\begin{eqnarray}
j(x,t) &=& \frac{e}{2 \pi \hbar } \sum_{X_1,X_2=R,L}
\int_{E_B}^{E_T} \int_{E_B}^{E_T} dE dE' e^{i (E-E') t/
\hbar} \nonumber \\
& & \hspace{1cm} \times A_{L}^{X_1 X_2}(E,E';x) \,
a^{\dagger}_{X_1 E}\, a^{}_{X_2 E'} . \label{JL-op}
\end{eqnarray}
Here, $E_B$ and $E_T$ denote the bottom and top values of the
energy band of the channel. The current matrix elements
$A_{L}^{X_1 X_2}$ are defined as
\begin{widetext}
\begin{eqnarray}
\displaystyle A_{L}^{L L} (E,E'; x) &=& \frac{1}{e}
\frac{\Omega}{\sqrt{v(E) v(E')}} \left( j_{L}^{\alpha E ; \alpha
E'}(x) + j_{L}^{\beta E ; \beta E'}(x) \, r^{*}(E) \, r^{}(E') \,
+j_{L}^{\alpha E ; \beta E'}(x) \, r^{}(E') +j_{L}^{\beta E ;
\alpha E'}(x)
\, r^{*}(E) \right) , \nonumber \\
\displaystyle A_{L}^{R R} (E,E'; x) &=& \frac{1}{e}
\frac{\Omega}{\sqrt{v(E)
v(E')}} j_{L}^{\beta E ; \beta E'}(x) \, t^{\prime *}(E) \, t^{\prime}(E') , \label{A^L-def} \\
\displaystyle A_{L}^{L R} (E,E'; x) &=& \frac{1}{e}
\frac{\Omega}{\sqrt{v(E) v(E')}} \left( j_{L}^{\beta E ; \beta
E'}(x) \, r^{*}(E) \, t^{\prime}(E') +
j_{L}^{\alpha E ; \beta E'}(x) \, t^{\prime}(E') \right) , \nonumber \\
\displaystyle A_{L}^{R L} (E,E'; x) &=& \frac{1}{e}
\frac{\Omega}{\sqrt{v(E) v(E')}} \left( j_{L}^{\beta E ; \beta
E'}(x) \, {t^\prime}^{*}(E) \, r^{}(E') + j_{L}^{\beta E ; \alpha
E'}(x) \, {t^{\prime}}^{*}(E) \right) , \nonumber
\end{eqnarray}
\end{widetext}
where $\Omega$ is the total length of the system, $v(E)=\hbar^{-1}
\partial E/\partial k$ the band velocity, and $j_{L}^{\alpha E ; \beta E'}(x)$
are the matrix elements of the appropriate current
operator\cite{Foot2} in the basis of the eigenfunctions $\alpha_{L
E}(x)$ and $\beta_{L E'}(x)$. For instance, for electrons with
parabolic dispersion relation
\begin{eqnarray}
&& j_{L}^{\gamma E ; \gamma' E'}(x) = \\
&& \frac{\hbar e}{2mi} \Bigl( \gamma^*_{L
  E}(x) \partial_x \gamma'_{L E'}(x) - (\partial_x \gamma^*_{L E}(x))
\gamma_{L E'}'(x) \Bigr) , \nonumber
\end{eqnarray}
where $\gamma, \gamma' = \alpha, \beta$. Inserting then
Eq.~(\ref{JL-op}) into Eq.~(\ref{noise}), and evaluating the
averages with respect to the incoming electron operators $a_{L E}$
and $a_{R E}$, the noise is easily determined. Notice that,
differently from the customary assumptions of linearized band and
energy independent transmission coefficients which lead to
Eq.~(\ref{shot-LB}), here Eq.~(\ref{A^L-def}) has been derived
without any specific hypothesis about the band dispersion and the
energy dependence of transmission
coefficients\cite{note-on-blanter00}.


\subsection{Voltage dependence of zero frequency noise}
\begin{figure}
\vspace{0.3cm}
\begin{center}
\epsfig{file=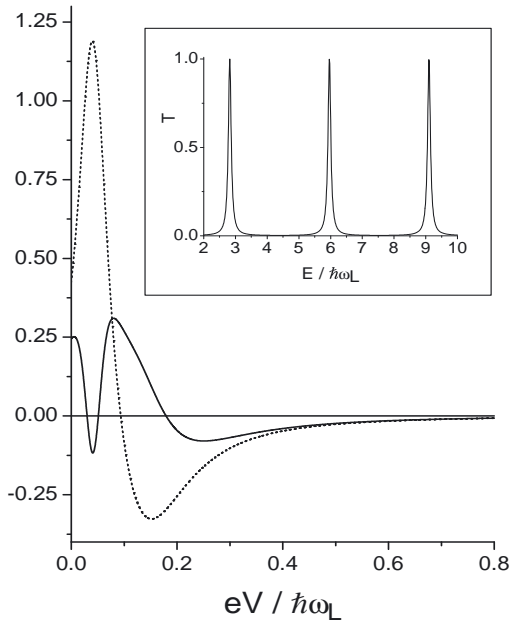,height=8.5cm,
width=7.5cm} \caption{\label{Fig2} The case of large capacitive
coupling $c/c_q \rightarrow \infty$: Voltage derivative of the
shot noise (\ref{shot-non-int}) in units of $e^3/2 \pi \hbar$ (solid curve)
and conductance (\ref{G-non-int}) in units of $e^2/2 \pi \hbar$ (dotted
curve) as a function of bias $V$ for a quantum wire with linear
dispersion relation and two impurities of equal strength
$\Lambda/\hbar v_F=3$, located at the contacts $x_{1,2}=\pm L/2$.
The electrochemical potentials are varied asymmetrically
$\mu_{L/R}=E_F + \gamma_{L/R} eV$, where $\gamma_L=3/2$ and
$\gamma_R=1/2$, and the Fermi level $E_F$ lies near the resonance
at $E_F=5.9 \hbar \omega_L$ with $\omega_L$ given by
Eq.~(\ref{omegaL}). At low bias $V \rightarrow 0$ the voltage
derivative of the noise and the conductance are positive (due to
the fact that $S(\omega=0;V=0)=0$, $S \ge 0$, and $I(V=0)=0$ with
$I \ge 0$ for $V \ge 0$). Negativity of both quantities is however
present for a wide range of voltage values up to $eV \simeq \hbar
v_F/L$, indicating that the current and the shot noise can
decrease with bias. Interestingly, a decreasing noise does not
necessarily correspond to a decreasing conductance. Inset: Energy
dependence of the transmission coefficient showing resonances.}
\end{center}
\end{figure}
To characterize the voltage dependence of the noise, we
investigate the derivative $\partial S/\partial V$. In this
section we consider the case $\omega=0$; since we assume the
temperature to be vanishing, no thermal noise sources are present,
and the noise at $\omega=0$ can be identified with the shot noise.
Denoting by $\mu_L$ and $\mu_R$ the electrochemical potentials of
the left and right leads, we assume for definiteness
$\mu_L-\mu_R=eV \ge 0$ and $E_B < \mu_L,\mu_R < E_T$. To be
general, we consider an arbitrary way to apply the bias $eV$, and
set $\mu_{L/R}=E_F+\gamma_{L/R} e V$, where $\gamma_L \ge 0$ is an
arbitrary coefficient and $\gamma_R=\gamma_L-1$. Note that this is
experimentally relevant in mesoscopic devices, where the bias is
not always applied symmetrically \cite{Leo}. We find that the
voltage derivative of the noise at zero frequency and zero
temperature is
\begin{eqnarray}
\frac{\partial S}{\partial V}(\omega=0,V) &=& \frac{e^3}{2 \pi \hbar}
  \Bigl[ \gamma_L \, T(\mu_L) R(\mu_L) - \nonumber \\
  & & \hspace{0.6cm} - \gamma_R T(\mu_R) R(\mu_R) \Bigl]
 \label{shot-non-int}
\end{eqnarray}
where $T(E)=|t(E)|^2$ and $R(E)=1-T(E)$ are the (energy-dependent)
transmission and reflection coefficients, respectively. As
mentioned above, the shot noise is independent of the point of
measurement $x$. The related expression for the differential
conductance $G(V)=dI(V)/dV$ reads
\begin{equation}
G(V)=\frac{e^2}{2 \pi \hbar} \left( \gamma_L T(\mu_L) -\gamma_R T(\mu_R)
\right) \label{G-non-int}
\end{equation} Our results Eq.~(\ref{shot-non-int}) and (\ref{G-non-int}) are quite general, since they hold for any dispersion relation, for electrons in the continuum, for Bloch electrons, for lattice models, and for
any number of scatterers. The only underlying hypothesis made is
that electrons in the wire do not experience any band-bottom shift
originating from the coupling to the gate~($c \gg c_q$). When the
transmission coefficient is energy independent, the noise
(conductance) is always increasing (constant) as a function of
bias, regardless of the values of $\gamma_L$ and $\gamma_R$, and
the results (\ref{shot-LB}) and (\ref{G-LB}) are thus recovered. A
qualitatively different behavior can occur if the energy
dependence is taken into account. To this purpose, it is
worthwhile discussing the role of the weights $\gamma_L$ and
$\gamma_R$. In the present case of very large capacitive coupling,
the transport properties of the wire depend in general not only on
the {\it difference} $eV$ between the electrochemical potentials
of the two reservoirs, but on $\mu_L$ and $\mu_R$ separately, and
thus on the specific values of $\gamma_L$ and $\gamma_R$. This
behavior is quite different from the case of weak capacitive
coupling, $c/c_q \rightarrow 0$, discussed in the next section
where the band bottom shift adjusts to the average value
$(\mu_L+\mu_R)/2$ in order to keep the wire electroneutral. In
that case, therefore, the values of $\gamma_R$ and $\gamma_L$ are
effectively fixed\cite{NOTE}, whereas in the case considered here
($c \gg c_q$) they can take any value and a richer scenario
arises.

For $\gamma_L \in [0,1]$, Eq.~(\ref{shot-non-int}) yields
$\partial S/\partial V \ge 0$ and the shot noise is an increasing
function of bias. In contrast, for $\gamma_L > 1$, one has
$\gamma_R \ge 0$, and the two terms in Eqs.~(\ref{shot-non-int})
and (\ref{G-non-int}) have competing signs. While for $V
\rightarrow 0$ the positive sign prevails (as expected from the
positivity of the noise and the linear conductance), the balance
can be different at finite bias, depending on the
energy-dependence of the transmission coefficient. This is
particularly interesting near a resonance, where slight variations
of $\mu_L$ and $\mu_R$ can yield large variations of the product
$T(E) R(E)$. In Fig.~\ref{Fig2} we show the voltage derivative of
the shot noise (solid curve) and the conductance (dotted curve) in
a double impurity wire with linear band dispersion. Typically, the
two impurities model the backscattering at non-ideal contacts. The
Fermi level lies near the second resonance peak shown in the inset
of Fig.~\ref{Fig2}, and $\gamma_L=3/2$. As one can see, the shot
noise has a non-linear behavior as a function of the bias, and
regimes of negativity in $dS/dV$ appear. A similar behavior is
exhibited by the conductance. Interestingly, when the solid curve
of Fig.~\ref{Fig2} undergoes a first dip, the dotted curve is
still positive, indicating that there exist voltage ranges in
which the current fluctuations decrease with bias in spite of the
fact that the average current increases. This is due to the fact
that, while the conductance is proportional to the transmission
coefficient $T$, the noise depends on the combination $T \, R$, so
that near a resonance the balance between the two terms
in Eq.~(\ref{shot-non-int}) can  differ  from the one in Eq.~(\ref{G-non-int}).\\
The voltage range in which the shot noise decreases with bias is
of order $\hbar \omega_L/e$, where
\begin{equation}
\omega_L \doteq v_F/L \quad , \label{omegaL}
\end{equation}
with $L$ being the distance between the two impurities. The energy
scale $\hbar \omega_L$ corresponds to the transversal time of the
wire of length $L$. For peaked resonances, like in
Fig.~\ref{Fig2}, $\hbar \omega_L$ can thus be interpreted as the
level spacing of the quantum dot defined by the two impurities.

While the voltage derivative of the $\omega=0$ shot noise can
become negative only if $\gamma_L >1$, at finite frequency this
condition is no longer necessary.

\subsection{Voltage dependence of finite frequency noise}
\label{Vdepn}

Since by now it has become experimentally relevant
\cite{schoe97,deblo03,Onac1,deblo06,Onac2,Glattli03} to understand
what happens at finite noise frequency $\omega$, we address this
regime in the following. At finite frequency, the scenario is much
richer, and a non-monotonic behavior of the noise may arise even
at zero temperature, and even independent of the condition
$\gamma_L > 1$ found for the shot noise. The generalization of
Eq.~(\ref{shot-non-int}) to finite frequencies can be readily
derived and reads
\begin{widetext}
\begin{eqnarray} \label{dSdV}
 \frac{\partial S}{\partial V}(x,\omega,V) &=&
 \frac{e^3}{2 \pi \hbar} \cdot \sum_{s=\pm} s \, \, \Bigr\{
\theta(\omega) \left( \gamma_L |A_{L}^{LL}(\mu_L,\mu_L+s\hbar \omega;x) |^2 \, + \, \gamma_R |A_{L}^{RR}(\mu_R,\mu_R+s\hbar \omega;x) |^2 \right) \, + \nonumber \\
&+& \theta (\hbar \omega + s \, eV) \left( \gamma_L
|A_{L}^{LR}(\mu_L,\mu_L+ s \hbar \omega;x) |^2 - \gamma_R
|A_{L}^{RL}(\mu_R,\mu_R- s \hbar \omega;x) |^2 \right) \Bigr\} ,
\end{eqnarray}
\end{widetext}
where $\theta(x)$ is the Heaviside function, and we have assumed
that $E_B < \mu \pm \hbar \omega,\mu_R \pm \hbar \omega < E_T$. In
Eq.~(\ref{dSdV}), the first and the second line describe
contributions to the voltage derivative caused by electrons
originating from the same lead and from different leads,
respectively. We see that Eq.~(\ref{dSdV}) contains terms with
competing signs. This originates from the energy dependence of the
expectation values \cite{blanter00}
\begin{eqnarray}
\langle a^{\dagger}_{X_1 E_1} a^{}_{X_2 E_2} a^{\dagger}_{X_3 E_3}
a^{}_{X_4 E_4} \rangle - \langle a^{\dagger}_{X_1 E_1} a^{}_{X_2
E_2} \rangle \langle a^{\dagger}_{X_3 E_3} a^{}_{X_4 E_4} \rangle
\nonumber
\end{eqnarray}
which determine the noise. Denoting by $f_{X_i}(E)$ the Fermi
distribution function of the electrons incoming from lead $X_i$,
one can easily see that at frequency $\omega$ the terms that
contribute to the above expectation values come form a ``Fermi
box'' $f_{X_1}(E) [ 1-f_{X_2}(E \pm \hbar \omega) ]$, the size of
which depends on the applied voltage \cite{Foot3}. An increase of
the electrochemical potential of lead $X_1$ ($X_2$) increases
(decreases) the noise strength. Only if the factors $|A^{X_1
X_2}_L|^2$ are energy-independent do these contributions cancel
out. This occurs, for instance, in a clean quantum wire with
linear band dispersion. In general, however, this is not the case.
An energy dependence of these factors can arise from resonance
phenomena, or from band curvature.

Let us consider the former contribution. When more than one
impurity are present, Fabry-P\'erot resonance phenomena occur
\cite{Peca,liang01} yielding a decrease of the noise with
increasing voltage even in the absence of band curvature
\cite{Recher}. We shall illustrate this effect with the simplest example
of a wire with linear dispersion relation and two impurities at
the positions $x_1$ and $x_2$ (e.g.\ at the non-ideal contacts).
The Hamiltonian of that problem is given by $H=H_0+H_{\rm imp}$,
where $H_0$ is given by Eq.~(\ref{H0}) and $H_{\rm imp}=H_{\rm
FS}+H_{\rm BS}$, with\cite{vonDelft,Herm00}
\begin{eqnarray}
H_{\rm FS} &=& \! \sum_{i=1,2} \Lambda^{\rm FS}_{i} \Bigl( :
\Psi^\dagger_R \Psi^{}_R : \, + \, : \Psi^\dagger_L
\Psi^{}_L : \Bigr)\Big|_{x=x_{i}} \label{HFS} \\
H_{\rm BS} &=& \! \sum_{i=1,2} \Lambda^{\rm BS}_{i} \Bigl( :
\Psi^\dagger_R \Psi^{}_L : \, + \, : \Psi^\dagger_L \Psi^{}_R :
\Bigr)\Big|_{x=x_{i}} \label{HBS}
\end{eqnarray}
where $\Psi_{R/L}(x)$ is the electron field operator for a
right/left-moving particle and $x_{i}$ is the position of the
$i$-th impurity. It turns out that the forward scattering term
$H_{\rm FS}$ just renormalizes the Fermi velocity
$v_F$\cite{vonDelft,Herm00}. In contrast, the backscattering term
$H_{\rm BS}$ is responsible for an energy-dependent transmission
through the system.

The scattering matrix of the system can be conveniently written in
terms of the elements $M_{nm}$ of the transfer matrix
$\mathsf{M}(E)$
\begin{equation}
\mathsf{S}(E) = \frac{1}{M_{22}(E)} \left(
\begin{array}{lr}
M_{12}(E) & 1
\\ 1 & -M^*_{21}(E)
\end{array} \right) \label{S-matr}
\end{equation}
with
\begin{eqnarray} \label{Ms}
M_{12} &=& -i \tilde{\Lambda}
\left[\left(1-i\tilde{\Lambda}\right)e^{-2ik x_1} +
\left(1+i\tilde{\Lambda}\right)e^{-2ik x_2} \right] , \nonumber \\
M_{22} &=& \left(1+i\tilde{\Lambda}\right)^2 +\tilde{\Lambda}^2
e^{-2ik(x_2-x_1)} ,
\end{eqnarray}
where $\tilde{\Lambda} \equiv \Lambda/(\hbar v_F)$ and we have
assumed (for simplicity) that $\Lambda \equiv \Lambda^{\rm BS}_{1}
= \Lambda^{\rm BS}_{2}$. The voltage derivative of the noise can
then be obtained by inserting the elements of the
$\mathsf{S}$-matrix into Eq.~(\ref{A^L-def}), and the latter into
Eq.~(\ref{dSdV}).
\begin{figure}
\vspace{0.3cm}
\begin{center}
\epsfig{file=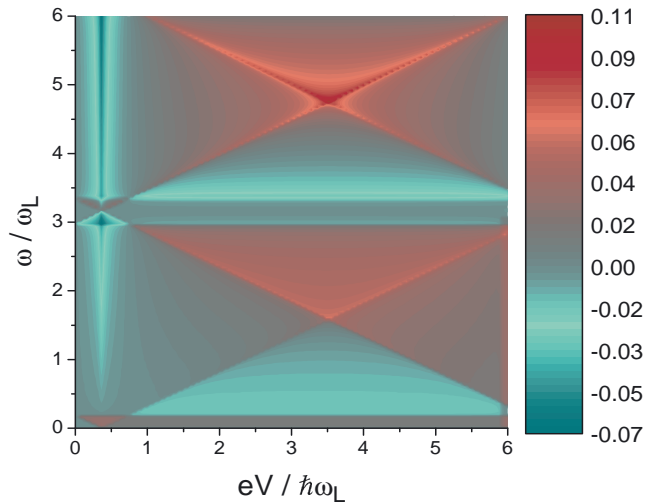,height=6.8cm,width=8.6cm}
\caption{\label{Fig3} The case of large capacitive coupling $c/c_q
\rightarrow \infty$: For a quantum wire with linear dispersion
relation and two impurities of equal strength $\Lambda/\hbar
v_F=10$ located at positions $x_{1,2}=\pm L/2$ the excess noise
(in units of the energy $\hbar \omega_L$ associated with the
ballistic frequency (\ref{omegaL})) is shown as a function of the
noise frequency $\omega$ and the applied voltage $V$ for a
measurement point at $x = - 0.6 L$. The values of the
electrochemical potentials in the leads are $\mu_{L/R} = E_F \pm
eV/2$ with $E_F=6 \hbar v_F/L$. Regions of negative excess noise
are clearly visible. This negativity originates from Fabry-P\'erot
resonance phenomena.}
\end{center}
\end{figure}

We remark that while in the single impurity case the absolute
values of the entries of $\mathsf{S}$ do not depend on the
impurity position, for two impurities these moduli depend on the
phase factor $\exp{(i k L)}$ related to the distance $L=x_2-x_1$
between the impurities. The transmission coefficient then exhibits
resonances as a function of the energy with a typical period
$\Delta E= \pi \hbar v_F/L$. In this case, even when the energy of
the excitations is within the range of validity of a linearized
band spectrum, the response of the system is non-linear and
quantum resonances emerge. For the average current these
Fabry-P\'erot resonances have been observed in a recent experiment
on carbon nanotubes \cite{liang01}. Here, we show that these
resonances also yield an oscillatory behavior of the voltage
derivative of the noise, and that in particular the excess noise
(\ref{SEX}) exhibits regions of negativity. Fig.~\ref{Fig3}
displays this effect for a wire with two impurities of equal
strength when the Fermi level is at resonance (corresponding to a
maximum of the transmission coefficient of the system). Note that
in the double impurity setup (similar to standard quantum dot
physics) the transmission depends on $\mu_L$ and $\mu_R$ and not
only on their difference $eV=\mu_L-\mu_R$. The reason is that the
energy landscape of the system (i.e.\ the energy levels in the dot
formed by the region between the two impurities) matters.
Fig.~\ref{Fig3} refers to the case $\gamma_L=-\gamma_R=1/2$; we
have checked, however, that the phenomenon of negative excess
noise in double impurity systems is rather generic and not just
related to a specific choice of $\gamma_L$ and $\gamma_R$.

\begin{figure}
\vspace{0.3cm}
\begin{center}
\epsfig{file=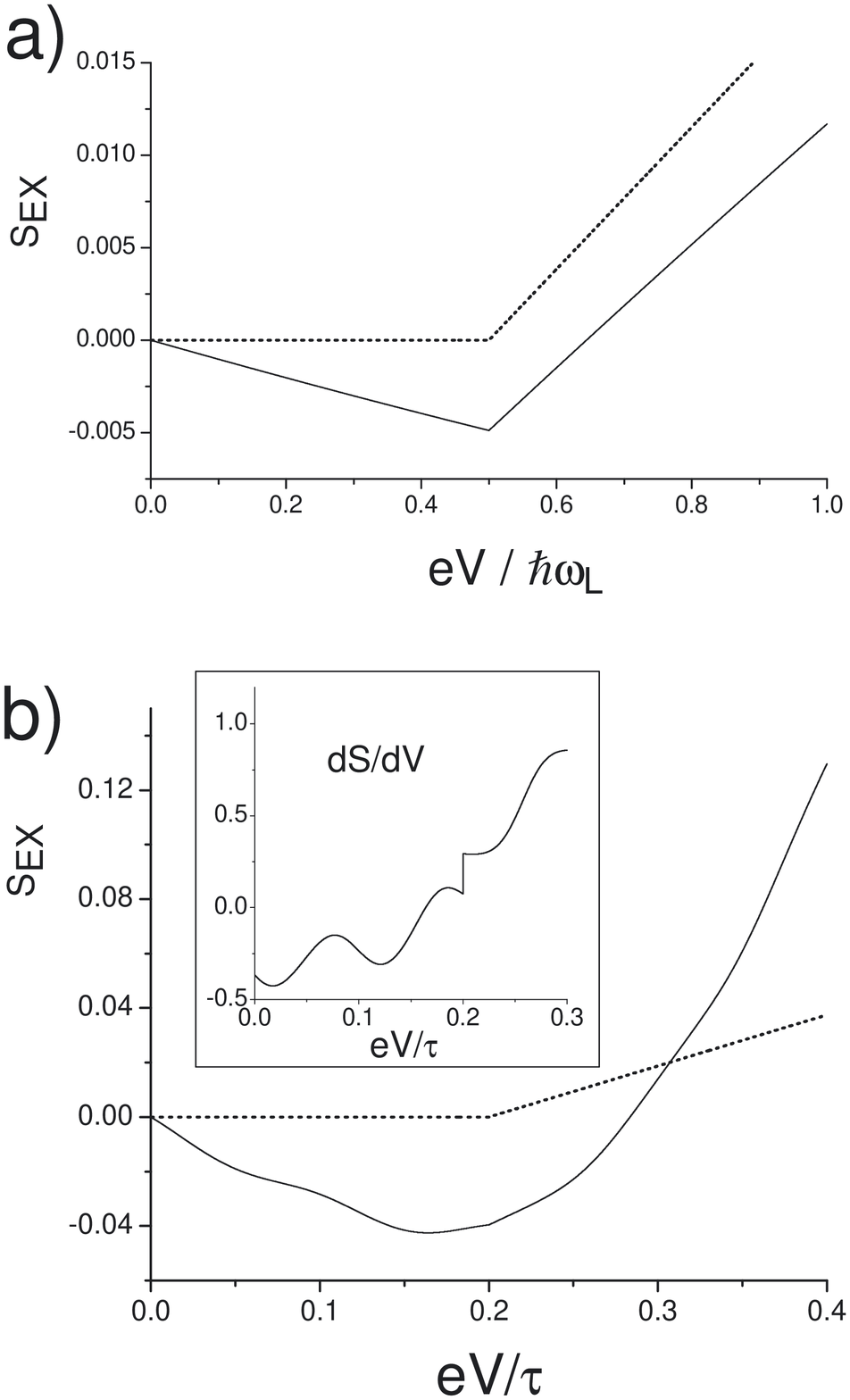,height=11cm,width=8.0cm}
\caption{\label{Fig4} The case of large capacitive coupling $c/c_q
\rightarrow \infty$: Negativity of the excess noise due to band
curvature effects for a quantum wire with one impurity.
\\a) Parabolic band curvature (\ref{disp-PAR}).
The solid curve shows the excess noise (in units of $e^2 \omega_L
/2\pi$ with $\omega_L$ given in Eq.(\ref{omegaL})) as a function
of the bias $V$ for a frequency $\omega=\omega_L/2$ at the
measurement point $x = - L$. The impurity of strength
$\Lambda/\hbar v_F=1$ is located in the middle; the Fermi level is
$E_F= 6 \hbar \omega_L$, and the bias weight is $\gamma_L=1$. The
dotted curve shows the excess noise for a wire with linear
dispersion and with the
same transmission coefficient at the Fermi level. \\
b) Sinusoidal band dispersion~(\ref{disp-TBM}) for the
tight-binding model~(\ref{H0TBM}). The solid curve shows the
excess noise (in units of $e^2 \tau /2 \pi \hbar $) as a function
of the bias $V$ for a frequency $\omega=0.2 \tau/\hbar$ at the
measurement point $j = - 51 a_0$ for a wire of length $L= 100
a_0$. The impurity of strength $\Lambda/a \tau=1$ is located in
the middle; the Fermi level is $E_F= \tau$, and the bias weight is
$\gamma_L=1$. The dotted curve has the same meaning as in a). The
slope $dS/dV$ is shown as an inset.}
\end{center}
\end{figure}

Let us now discuss the effect of band curvature. For simplicity we
assume that the wire has just one impurity, in order to rule out
contributions coming from resonance phenomena described above. We
recall that for a wire with linear dispersion the excess noise is
always non-negative\cite{blanter00}. In particular, $S_{\rm EX}=0$
for $eV < \hbar \omega$ and $S_{\rm EX} = (e^2/2 \pi \hbar) (eV-\hbar \omega)
T(1-T)$ for $eV > \hbar \omega$, with the well known singularity
at $\hbar \omega = eV $. \cite{Leso} In presence of band
curvature, this result is modified, as illustrated in
Fig.~\ref{Fig4}. In particular, Fig.~\ref{Fig4}a shows the case of
a wire with parabolic dispersion
\begin{equation}
E_k=\frac{\hbar^2 k^2}{2 m} , \label{disp-PAR}
\end{equation}
whereas Fig.~\ref{Fig4}b depicts the case of a lattice
tight-binding model~(\ref{H0TBM}) with dispersion
\begin{equation}
E_k=-4 \tau \cos{k a_0} \label{disp-TBM}
\end{equation}
The plots depict the noise as a function of bias at a fixed value
of the frequency $\omega$. As one can see, the excess noise is
negative and has a minimum around $eV \sim \hbar \omega$. For
comparison, we also display by a dotted line the excess noise for
a model with linear spectrum and with the same transmission
coefficient at the Fermi level. The singularity at $eV = \hbar
\omega$ is still evident in Fig.~\ref{Fig4}a, whereas in
Fig.~\ref{Fig4}b it is masked by the appearance of oscillations;
in this case, the inset showing the derivative $dS/dV$ helps to
locate the singularity. The oscillations are of the type discussed
in Ref.~\onlinecite{Trauz-Gra} and are related to the distance
between the impurity and the measurement point of the noise (here
chosen in the left lead, close to the contact). The fact that they
appear in Fig.~\ref{Fig4}b and not in Fig.~\ref{Fig4}a is merely
due to a difference in energy scales. For a lattice model energies
are indeed more naturally expressed in terms of the hopping
energy~$\tau$ rather than the energy $\hbar \omega_L$ related to
the length of the wire. In terms of a parabolic model, the case
shown in Fig.~\ref{Fig4}b corresponds to much higher energies than
in Fig.~\ref{Fig4}a and an oscillatory behavior becomes visible.
These oscillations are due to the backscattering contributions
$e^{\pm 2 i k (x-x_0)}$ to the electron density (Friedel
oscillations), where $x_0$ is the impurity position, $x$ the
measurement point, and $k$ the electron momentum, depending on the
energy through Eq.~(\ref{disp-PAR}).

\subsection{Experimental feasibility}

As can be seen from Fig.~\ref{Fig3}, provided the bias is
sufficiently high $eV \simeq \hbar \omega_L$, negative excess
noise is present already at relatively low noise frequencies
$\omega \lesssim 0.5 \omega_L$. The effect, however, is
particularly pronounced at frequencies $\omega$ of order $\pi
\omega_L$ or higher. In mesoscopic devices of dimension $L \approx
1 - 10 \mu {\rm m}$, like quantum wires based on seminconductor
heterostructures\cite{CEO} or single-wall carbon
nantotubes\cite{liang01,SWNT}, this corresponds to frequencies of
the order of 100 GHz to a few THz (for a typical Fermi velocity
$v_F$ of order 10$^6$ m/s). Such high frequencies require a
sophisticated measurement technique as the one proposed in
Ref.~\onlinecite{Aguado} and developed in
Ref.~\onlinecite{deblo03}. There, a detector (based on a SIS
junction) and a device are capacitively coupled on chip. This
allows noise detection over a large bandwidth (up to 100 GHz for
Al as superconductor and up to 1 THz for Nb as superconductor).
Thus, it should, in principle, be possible to observe regions of
negative excess noise.

As far as the region of validity of the non-interacting theory in
Sec.~\ref{Vdepn} is concerned, this depends on the magnitude of
the geometric capacitance $c$ compared with the quantum
capacitance~$c_q$. In principle, one can create mesoscopic devices
(where the conductor is, for instance, a carbon nanotube) both in
the interacting or the non-interacting regime, where~$c_q$ is much
larger or much smaller than~$c$, respectively \cite{Ilani}. One
way how the ratio $c/c_q$ can be controlled is by changing the
dielectric constant of the substrate, which changes only~$c$ and
not~$c_q$. Apart from the dielectric constant of the medium, $c$
depends on the shape and dimensions of the sample as well as on
its distance to the conducting plane.

\section{The case of finite capacitance}
\label{Sec_Lut}

We now discuss the effects of a finite capacitive
coupling~(\ref{HC}) between the wire of length~$L$ and the gate.
As mentioned in the introduction, a finite capacitance induces a
finite (frequency-dependent) band-bottom shift in the wire, which
in turn affects the current in the wire yielding essentially an
interacting transport problem. One way to face this problem is to
generalize the scattering matrix formalism: the scattering matrix~$\mathsf{S}$ 
depends in fact not only on the electron energy~$E$,
but also on the potential profile $U$, so that the coefficients~$A^{X_1 X_2}$ 
in Eq.~(\ref{A^L-def}) have to be modified
accordingly $A^{X_1 X_2} \rightarrow A^{X_1 X_2}+ \Delta A^{X_1
X_2}$. This approach has been introduced by B\"uttiker and
co-workers in a series of seminal papers
\cite{Chris-Butt,Butt-PLA-93,Butt-1,Butt-2,Butt-3}, and was
recently applied to the description of quantum point contacts and
chaotic cavities \cite{Pedersen,Hekk}. An alternative way is to
determine self-consistently the electron density taking into
account both the bare charge injected by the leads and the charge
induced by the fluctuations of the potential $U$ through the
polarization propagator $\Pi(x,x^{\prime})$. This technique has
been successfully applied\cite{Bla-But-EPL,Blanter} to the case of
a clean multi-channel quantum wire. In presence of impurities,
however, this method is practicable only for weak potential $U$,
and is therefore not suitable to investigate the regime of small
capacitance.

Here we shall thus adopt another approach which allows us to
explore the full range of $c/c_q$, and to account for impurities
as well. We shall exploit a mapping to a Luttinger liquid. It is,
in fact, well known \cite{Blanter,Glazman,Egger,SafiEPJB} that in
case that band curvature effects can be neglected, the problem of
a quantum wire with a geometric capacitance $c$ with respect to a
gate can be mapped onto a Luttinger liquid with interaction
strength
\begin{equation}
g=\frac{1}{\sqrt{1+\frac{c_q}{c}}} . \label{g-capacitance}
\end{equation}
where $c_q=e^2/\pi \hbar v_F$ is the quantum capacitance. The case
of large geometrical capacitance $c \gg c_q$ treated in the
previous section corresponds to $g=1$ (no charge screening),
whereas the opposite limit of $c \ll c_q$, i.e.\ $g \to 0$,
corresponds to a completely electroneutral wire (full charge
screening). In any intermediate case, only a fraction $1-g^2$ of
the charge is screened \cite{Egger}.  Notice also that the
parameter $g$ is strictly related to the electro-chemical
capacitance\cite{Butt-PLA-93} per unit length
\begin{equation}
\frac{1}{c_\mu}=\frac{1}{c}+\frac{1}{c_q}
\end{equation}
and to the charge-relaxation time\cite{Pedersen}
\begin{equation}
t_\mu=\frac{h}{e^2} c_\mu L
\end{equation}
through the relation
\begin{equation}
c_\mu=g^2 \, \frac{e^2}{\pi \hbar v_F}
\end{equation}
The charge-relaxation time is therefore determined by the
timescale of the wire uncoupled to the gate (typically the
ballistic time $L/v_F$) in the limit of $c/c_q \gg 1$, and by the
RC-time $t_{RC}=(h/e^2) c L$ in the limit of $c/c_q \ll 1$. This is
in accordance with the results of a recent
investigation\cite{Hekk} of the noise in mesoscopic chaotic
cavities coupled to a gate. Typically, it is the charge-relaxation
time that determines the dynamical time scales of transport in a
mesoscopic conductors\cite{Foot-RC}.

The mapping to the Luttinger liquid model is realized through the
standard bosonized representation \cite{gogol98} of electron field
operators
\begin{eqnarray}
\Psi_{R/L} (x) = \frac{\eta_{R/L}}{\sqrt{2 \pi a}} \, e^{\pm i
\sqrt{4 \pi} \Phi_{R/L}(x)}
\end{eqnarray}
where $a$ is the cut-off length, $\eta_{R/L}$ are Majorana
fermions (Klein factors), and $\Phi_{R/L}$ non-local plasmonic
excitation fields. Introducing the usual Bose field operator
$\Phi=\Phi_R+\Phi_L$ and its conjugate momentum $\Pi=\partial_x
(\Phi_R-\Phi_L)$, the Hamiltonian reads $H=H_{\rm LL} + H_{\rm
imp}$ with
\begin{eqnarray}
 H_{\rm LL} &=&\frac{\hbar v_F}{2} \int_{-\infty}^{\infty}
 dx \left[ \Pi^2 + \frac{1}{g^2(x)}
(\partial _x \Phi )^2 \right] , \label{HLL} \\
H_{\rm imp} &=& \sum_{i} \lambda_{i} \cos{[\sqrt{4 \pi}
\Phi(x_i)+2 k_F x_i]} . \label{HI}
\end{eqnarray}
The first term $H_{\rm LL}$ describes the band kinetics of the
quantum wire (\ref{H0}) as well as its capacitive
coupling~(\ref{HC}) to the gate. Note that the Luttinger liquid
parameter~(\ref{g-capacitance}) in the Hamiltonian~(\ref{HLL}) is
inhomogeneous. Indeed, although the capacitive coupling is finite
($0 < g < 1$) over the length $L$ of the wire (see
Fig.~\ref{Fig1}), we have an effectively large geometrical
capacitance ($g=1$) in the regions of the leads. This is because,
as far as transport properties of the wire are concerned, the
leads can be modelled as ideal Fermi gases with electrochemical
potentials $\mu_R$ and $\mu_L$ which determine the energy of the
electrons {\it injected} into the gated wire. A step-like
approximation for $g(x)$ is valid when the change of the
capacitive coupling at the ends of the wire occurs over a distance
much larger than the Fermi wavelength and much smaller than the
length of the wire. The second term (\ref{HI}) of $H$ represents
the backscattering by impurities at positions~$x_i$, which
introduces a strong non-linearity in the field $\Phi$. The
bosonized expression for the current electron operator
reads\cite{gogol98}
\begin{equation}
j(x) = \frac{e v_F}{\sqrt{\pi}} \,  \Pi(x) , \label{j-bos}
\end{equation}
and the noise can be computed by inserting Eq.~(\ref{j-bos}) into
Eq.~(\ref{noise}), yielding
\begin{equation}
S(x,\omega,V)=S_0(x,\omega)+S_{\rm imp}(x,\omega,V)
\label{S0+Simp}
\end{equation}
In the above equation, the first term describes the noise in the
absence of impurities, which can be computed exactly since the
Hamiltonian (\ref{HLL}) is quadratic. The second term accounts for
the effect of the impurities~(\ref{HI}), which can be treated
perturbatively within the non-equilibrium Keldysh
formalism\cite{keldysh}.

We recall that, in the absence of impurities, due to the linear
spectrum of the Hamiltonian, the noise~$S_0$ does not depend on
the voltage\cite{chamon96}, and therefore the excess noise for a
clean wire is vanishing. On the other hand, in presence of more
than one impurity, Fabry-P\'erot resonance phenomena arise, which
for $c/c_q \rightarrow \infty$ were shown in Sec.~II to lead to
negative excess noise. Not surprisingly, this behavior survives
also for finite capacitive coupling, as recent results for
Luttinger liquids at high frequencies show (see Fig.~7c and
Fig.~9a of Ref.~\onlinecite{Recher}).

The case of a single impurity in an adiabatically contacted wire
is therefore peculiar since Fabry-P\'erot interferences are
absent. This is the simplest example in which a finite capacitance
can yield notable and clearly identifiable differences with
respect to the case of large capacitive coupling, and we shall
thus focus on this situation.

\subsection{Voltage dependence of zero frequency noise}

In a 1D wire with one impurity a capacitive coupling~(\ref{HC}) to
a gate has dramatic effects on the transport properties. The
interplay of Friedel oscillations with density-density
correlations leads to a strong renormalization of the impurity
strength by the coupling to the gate, driving the wire at $T=0$
into an insulating state \cite{KaneFisher}. Electron transport
only occurs at finite bias, and in particular, for $eV \gg
\lambda^{*}$, where $\lambda^*=\hbar \omega_c (\lambda/\hbar
\omega_c)^{1/(1-g)}$ is the renormalized impurity
strength\cite{vonDelft,KaneFisher,PRBlong} at bandwidth $\hbar
\omega_c$. A similar behavior occurs in a one-channel coherent
conductor in series with an ohmic environment\cite{Safi-Saleur}.

Thus, in presence of a capacitive coupling, the voltage dependence
of the noise is not trivial: even at zero frequency (shot noise),
it can significantly deviate from the result (\ref{shot-LB})
obtained for $c/c_q \rightarrow \infty$. The method how to
calculate noise properties within the bosonization formalism has
been expounded in Ref.~\onlinecite{PRBlong}. Here we use these
earlier results to elucidate the effects of finite capacitive
coupling. The shot noise of a gated quantum wire in the weak
backscattering limit is proportional to the backscattering current
\cite{kane94}
\begin{equation}
S(\omega = 0,V) = e \, I_{\rm BS}(V) , \label{shot-LL}
\end{equation}
Here $I_{\rm BS}$ is the part of the transmitted current $I$ that
is backscattered by the impurities; explicitly
\begin{equation}
I = \frac{e^2}{2 \pi \hbar} V - I_{\rm BS}  \label{I-decomp}
\end{equation}
where the first term represents the current for the wire without
impurities. Eq.~(\ref{shot-LL}) has the same form as the relation
between shot noise and backscattering current obtained for
infinite capacitive coupling $c/c_q \rightarrow \infty$ in
presence of a weak impurity. In that case, indeed, the shot noise
is just given by Eq.~(\ref{shot-LB}) for the case of a single
channel with $T \simeq 1$, whereas $I_{\rm BS} = (e^2/2 \pi \hbar) V (1-T)$,
as easily obtained from Eqs.~(\ref{I-decomp}) and (\ref{G-LB}).
Here, however, the finite capacitive coupling entails a dependence
of $I_{\rm BS}$ on the bias voltage $V$ that differs significantly
from a mere proportionality, the latter being only valid in the
regime $c/c_q \rightarrow \infty$. Now, one rather
obtains\cite{PRL03} \begin{equation} I_{\rm BS}(V) =
I^{\infty}_{\rm BS}(V) \left[1+ f_{\rm BS} ( eV/\hbar \omega_L)
\right]. \label{IBS-fact}
\end{equation}
In Eq.~(\ref{IBS-fact}), the first factor shows power law behavior
as a function of the bias\cite{KaneFisher}
\begin{equation}
I^{\infty}_{\rm BS}(V) = \frac{e^2}{2 \pi \hbar} V \frac{\pi^2}{\Gamma(2g)}
\left( \frac{\lambda^*}{eV} \right)^{2 (1-g)} ,
\end{equation}
and the corresponding exponent is directly related to the
capacitive coupling through the parameter~(\ref{g-capacitance}).
The second factor  describes oscillatory deviations from
$I^{\infty}_{\rm BS}(V)$, related to the finite length of the wire
through the ballistic frequency (\ref{omegaL}).  Details about the
explicit form of $f_{\rm BS}$ can be found in
Ref.~\onlinecite{PRBlong}. Here we just mention that $f_{\rm BS}$
decays for $eV \gg \hbar \omega_L$. The general trend of the shot
noise as a function of the applied bias is thus determined by
$I^{\infty}_{\rm BS}(V)$, and one  obtains
\begin{equation}
\frac{\partial S}{\partial V}(\omega=0,V) \simeq (2g-1) \,
\frac{e^3}{\hbar} \frac{\pi}{ \Gamma(2g)}\, \left(
\frac{\lambda^{*}}{eV} \right)^{2(1-g)} ,
\end{equation}
For $1/2 < g <1$, i.e. for $c > c_q/3$, the voltage derivative is
positive and, in particular, in the limit $g \rightarrow 1$, the
result (\ref{shot-LB}) for large capacitance (in the special case
of weak impurity backscattering) is recovered. In contrast, for
\begin{equation}
c < \frac{c_q}{3}
\end{equation}
the shot noise {\it decreases} with increasing voltage.

\subsection{Voltage dependence of finite frequency noise}

\begin{figure}
\vspace{0.3cm}
\begin{center}
\epsfig{file=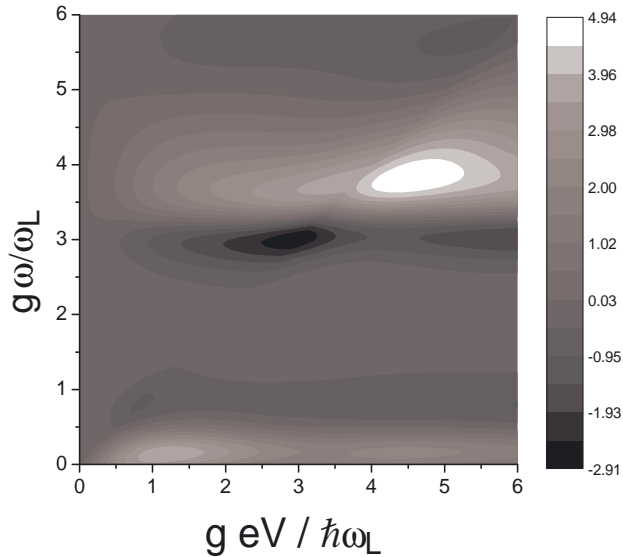,height=8.0cm,width=9.0cm}
\caption{\label{Fig5} Excess noise for a gated wire with
capacitance $c=c_q/10$, with an impurity shifted by $L/4$ off the
center of the wire. The measurement point is at $x = - 0.6 L$.
Regions of negative excess noise are visible at frequencies
$\omega \sim \pi \omega_L /g$, where $\omega_L$ is the ballistic
frequency (\ref{omegaL}) and $g$ is the parameter defined in
Eq.~(\ref{g-capacitance}). }
\end{center}
\end{figure}

At finite frequency, the differences with respect to the case $c
\gg c_q$ are even more striking. We recall that in the latter
regime the excess noise of a quantum wire with linear dispersion
and one (weak) impurity is $S_{\rm EX}=0$ for $eV < \hbar \omega$
and $S_{\rm EX} = (e^2/2 \pi \hbar) (eV-\hbar \omega) R$ for $eV > \hbar
\omega$, $R$ being the reflection coefficient. While there are no
Fabry-P\'erot resonances for a single impurity, when $c/ c_q$ is
finite, another type of resonance phenomena occurs due to the
difference in capacitive coupling strength between the gated
region and the metallic lead region. Indeed, when an interacting
1D wire is connected to metallic leads backscattering can occur at
adiabatic contacts due to Andreev-like reflections of plasmonic
charge excitations
\cite{safi95}. Finite length effects thus emerge. \\
Using Eq.~(\ref{S0+Simp}) and recalling the fact that the term
$S_0$ does not depend on the bias voltage $V$, the excess noise
can be written as
\begin{equation}
S_{\rm EX}(x,\omega,V)=S_{\rm imp}(x,\omega,V)-S_{\rm
imp}(x,\omega,V=0)
\end{equation}
Here the term $S_{\rm imp}(x,\omega,V)$ can be shown to be a
product of a local current correlator, which depends on the
voltage $V$, evaluated at the impurity position $x_0$ and the
clean-wire non-local retarded correlator between the measurement
point $x$ and the impurity position $x_0$ (see
Ref.~\onlinecite{PRBlong} for details). The Andreev-type
reflections enter directly in these correlators, so that the
interference phenomena between Andreev-type reflected plasmonic
charge excitations and modes backscattered at the impurity give
rise to notable structure in the voltage-frequency diagram of the
excess noise, as illustrated in Fig.~\ref{Fig5} for the case of a
small capacitance $c = c_q/10$. Regions of negative excess noise
are present at frequencies of order $\omega_L/g$. In
Fig.~\ref{Fig5} the noise is taken in its symmetrized version, and
the impurity is off-centered at $x_0=L/4$; results for the more
special case of a centered impurity can be found in
Ref.~\onlinecite{PRBlong}.

\section{Conclusions}
\label{Sec_con}

We have shown that in the quantum regime the current noise of a
quantum wire capacitively coupled to a gate can decrease with
increasing applied voltage. Both the shot noise and the finite
frequency noise have been analyzed in different regimes, ranging
from large to small ratios $c/c_q$, where $c$ is the geometrical
and $c_q=e^2 \nu$ the quantum capacitance. In particular, in the
regime of large capacitive coupling $c \gg c_q$, the noise can
decrease with increasing bias voltage when the transmission
coefficients are energy dependent. This can occur either in the
presence of Fabry-P\'erot resonances, like for instance in a
non-ideally contacted wire, or because of band curvature effects.
In both cases we have shown that excess noise can be negative.

When the coupling to the gate is of order $c_q$, even for energy
independent transmission coefficient and in the absence of
Fabry-P\'erot resonances or band curvature, the noise exhibits
rich structure which can significantly deviate from the usual
expression (\ref{shot-LB}). For instance, for an adiabatically
contacted wire with one impurity, when the geometrical capacitance
is smaller than~$c_q/3$, the shot noise decreases with bias as a
power law. At finite frequency, Andreev-type resonances due to the
finite length of the wire induce negativity of the excess noise.
We have thus demonstrated that the quantity usually termed {\it
excess} noise can become negative even at zero temperature. Furthermore, we
have estimated that such negative excess noise can be observed
with state-of-the-art techniques using on-chip noise detection
schemes \cite{Aguado,deblo03}. We therefore expect that these
predictions can be verified in current experimental realizations
of ballistic 1D conductors, such as quantum wires based on {\rm
GaAs/AlGaAs} heterostructures\cite{CEO}, or single-wall carbon
nanotubes\cite{liang01,SWNT}.

\acknowledgments

We would like to thank H.~Bouchiat, M.~B{\"u}ttiker, R.~Deblock,
S.~Ilani, and L.P.~Kouwenhoven for interesting discussions. This
work was supported by the EU network DIENOW and the Dutch Science
Foundation NWO/FOM.


\begin{thebibliography}{99}

\bibitem{CW} H.B. Callen and T.A. Welton, Phys. Rev. {\bf 83},
34 (1951).

\bibitem{Leso-Loos-JETP97} G.B. Lesovik, and R. Loosen, JETP Lett. {\bf 65}, 295 (1997).

\bibitem{Landau} L.D. Landau and E.M. Lifshitz, {\it Statistical Physics Part 1}, 3rd ed.
(Butterworth Heinemann, 1997).

\bibitem{Kogan} S. Kogan, {\it Electronic Noise and Fluctuations in Solids} (Cambridge University Press, 1996).



\bibitem{Aguado}
R. Aguado and L.P. Kouwenhoven, Phys. Rev. Lett. {\bf 84}, 1986
(2000).

\bibitem{deblo03}
R. Deblock, E. Onac, L. Gurevich, and L.P. Kouwenhoven, Science
{\bf 301}, 203 (2003).

\bibitem{blanter00}
Y.M. Blanter and M. B{\"u}ttiker, Phys. Rep. {\bf 336}, 1 (2000).


\bibitem{Trauz-Gra} B. Trauzettel and H. Grabert, Phys. Rev. B, {\bf 67}, 245101 (2003).

\bibitem{Khlus87} V. A. Khlus, Sov. Phys. JETP {\bf 66}, 1243 (1987).

\bibitem{Leso89} G.B. Lesovik, JETP Lett. {\bf 49}, 592 (1989).

\bibitem{MBPRL90} M. B\"uttiker, Phys. Rev. Lett. {\bf 65}, 2901 (1990).

\bibitem{trauz04}
B. Trauzettel, I. Safi, F. Dolcini, and H. Grabert, Phys. Rev.
Lett. {\bf 92}, 226405 (2004).

\bibitem{engel04}
H.-A. Engel and D. Loss, Phys. Rev. Lett. {\bf 93}, 136602 (2004).

\bibitem{lebed04}
A.V. Lebedev, A. Cr{\'e}pieux, and T. Martin, Phys. Rev. B {\bf
71}, 075416 (2005).

\bibitem{Pham} K.-V. Pham, cond-mat/0504389 (unpublished).

\bibitem{Pisto} D. Bagrets and F. Pistolesi, cond-mat/0606775 (unpublished).

\bibitem{schoe97}
{R.J. Schoelkopf}, P.J. Burke, A.A. Kozhevnikov, D.E. Prober, and
M.J. Rooks, Phys. Rev. Lett. {\bf 78}, 3370 (1997).

\bibitem{Onac1}
E. Onac, F. Balestro, B. Trauzettel, C.F.J. Lodewijk, and L.P.
Kouwenhoven, Phys. Rev. Lett. {\bf 96}, 026803 (2006).

\bibitem{deblo06}
P.-M. Billangeon, F. Pierre, H. Bouchiat, and R. Deblock, Phys.
Rev. Lett. {\bf 96}, 136804 (2006).

\bibitem{Onac2}
E. Onac, F. Balestro, L.H. Willems~van~Beveren, U.~Hartmann, Y.V.
Nazarov, and L.P. Kouwenhoven, Phys. Rev. Lett. {\bf 96}, 176601
(2006).

\bibitem{Leso-Loos} G.B. Lesovik and R. Loosen, Z. Phys. B {\bf
91}, 531 (1993).

\bibitem{Bula-Rubi} O.M. Bulashenko and J.M. Rub{\`i}, Physica E
{\bf 17}, 638 (2003).

\bibitem{WK} N. Wiener, Acta Math. {\bf 55}, 117 (1930); A. Khintchine, Math. Ann. {\bf 109}, 604 (1934).

\bibitem{Blanter}
Ya.M. Blanter, F.W.J. Hekking, and M. B{\"u}ttiker, Phys. Rev.
Lett. {\bf 81}, 1925 (1998).

\bibitem{SafiEPJB} I. Safi, Eur. Phys. J. B {\bf 12}, 451 (1999).

\bibitem{Foot1} The coupling to the gate has negligible effects on the metallic leads, due to their electroneutrality, and is therefore assumed to be zero here.
\bibitem{vonDelft} J. von Delft, and H Sch{\"o}ller, Ann. Phys. (Leipzig), {\bf 4}, 225 (1998).


\bibitem{Foot2} Regarding the appropriate current operator we refer to Eq.~(\ref{j-PAR}) for a parabolic dispersion band, to Eq.~(\ref{j-DIR}) for a Dirac linearized band, and to Eq.~(\ref{j-TBM}) for a lattice tight-binding model.

\bibitem{note-on-blanter00} Similar equations for the coefficients (\ref{A^L-def}) have been obtained in Ref.~\onlinecite{blanter00}; there, however, the assumptions of a linearized band and energy independent transmission were immediately made in order to simplify the treatment.

\bibitem{Leo}
L.P. Kouwenhoven, private communication.

\bibitem{NOTE} For an energy-independent transmission coefficient,
for instance, one effectively has $\mu_{L/R}=E_F \pm eV/2$. In
other situations, the determination of the band-bottom shift is
more involved. For resonant tunneling see e.g.
Ref.~\onlinecite{Chris-Butt}.

\bibitem{Glattli03} L.-H. Reydellet, P.~Roche, D.C.~Glattli, B.~Etienne, and Y.~Jin, Phys. Rev. Lett. {\bf 90}, 176803 (2003).


\bibitem{Foot3}
The expression ``Fermi box'' seems to be evocative here, since the
expression $f_{X_1}(E) [1-f_{X_2} (E \pm \hbar \omega)]$  as a
function of energy $E$ has the shape of a box. At zero temperature
it is indeed 1 for $\mu_{X_1} \le E \le \mu_{X_2} \mp \hbar
\omega$, and 0 elsewhere.
\bibitem{Peca}
C.S. Peca, L. Balents, and K.J. Wiese, Phys. Rev. B {\bf 68},
205423 (2003).

\bibitem{liang01}
W. Liang, M. Bockrath, D. Bozovic, J.H. Hafner, M. Tinkham, and H.
Park, Nature {\bf 411}, 665 (2001).


\bibitem{Recher} P. Recher, N.Y. Kim, and Y. Yamamoto, cond-mat/0604613
(unpublished).

\bibitem{Herm00} See Sec.~3.3 of H. Grabert in ``Exotic States in Quantum Nanostructures'' ed. by S. Sarkar, Kluwer (2002); cond-mat/0107175.




\bibitem{Leso} G.B. Lesovik, JETP Lett. {\bf 70}, 208 (1999).


\bibitem{CEO} R. de Picciotto, H. L. Stormer, L.~N. Pfeiffer, K.~W. Baldwin, and K.~W.
West, Nature {\bf 411}, 51 (2001); A. Yacoby, H.~L. Stormer, N. S.
Wingreen,  L.~N. Pfeiffer, K.~W. Baldwin, and K.~W. West, Phys.
Rev. Lett. {\bf 77},  4612  (1996); Solid State Comm. {\bf 101},
77  (1997); S. Tarucha, T. Honda, and T. Saku, Solid State Comm.
{\bf 94},  413  (1995).

\bibitem{SWNT} M. Bockrath, D.~H. Cobden, J. Lu, A.~G. Rinzler, R.~E. Smalley,
L. Balents, and P.~L. McEuen, Nature {\bf 397},  598  (1999); C.
T. White, and T. N. Todorov, Nature {\bf 393}, 240 (1998); C.
Dekker, Phys. Today {\bf 52}, 22 (1999); S. Tans. M. H. Devoret,
H. Dai, A. Thess, R. E, Smalley, L.J. Geeligs, and C. Dekker,
Nature {\bf 386}, 474 (1997);


\bibitem{Ilani}
S. Ilani, L.A.K. Donev, M. Kindermann, and P.L. McEuen, Nature
Phys. {\bf 2}, 687 (2006); S. Ilani, private communication.

\bibitem{Chris-Butt} T. Christen and M. B{\"u}ttiker, Europhys. Lett.
{\bf 35}, 523 (1996).

\bibitem{Butt-PLA-93} M. B{\"u}ttiker, H. Thomas, and A. Pr{\^e}tre, Phys. Lett. {\bf
180A}, 364 (1993).

\bibitem{Butt-1}
M. B{\"u}ttiker, A. Pr{\^e}tre, and H. Thomas, Phys. Rev. Lett.
{\bf 70}, 4114 (1993); M. B{\"u}ttiker, J. Phys.: Condens. Matter
{\bf 5}, 9361 (1993).

\bibitem{Butt-2}
P.W. Brouwer and M. B{\"u}ttiker, Europhys. Lett. {\bf 37}, 441
(1997).

\bibitem{Butt-3} K.E. Nagaev, S. Pilgram, and M.
B{\"u}ttiker, Phys. Rev. Lett. {\bf 92}, 176804 (2004).

\bibitem{Pedersen}
M.H. Pedersen, S.A. van Langen, and M. B{\"u}ttiker, Phys. Rev. B
{\bf 57}, 1838 (1998).

\bibitem{Hekk}
F.W.J. Hekking and J.P. Pekola, Phys. Rev. Lett. {\bf 96}, 056603
(2006).

\bibitem{Bla-But-EPL} Ya. M. Blanter and M. B\"uttiker, Europhys. Lett. {\bf 42}, 535 (1998).




\bibitem{Glazman}
L.I. Glazman, I.M. Ruzin, and B.I. Shklovskii, Phys. Rev. B {\bf
45}, 8454 (1992).

\bibitem{Egger}
R. Egger and H. Grabert, Phys. Rev. Lett. {\bf 79}, 3463 (1997).

\bibitem{Foot-RC} For chaotic cavities in the regime  $c \ll c_q$ there
exist exceptions to this general statement.
In situations where weak localization effects become relevant, for instance,
the AC conductance fluctuations  are governed by the dwell time (see Ref.~\onlinecite{Butt-2}).
Similarly, the latter also enters in the third cumulant of the electrical current (see Ref.~\onlinecite{Butt-3}).


\bibitem{gogol98}
A.O. Gogolin, A.A. Nersesyan, and A.M. Tsvelik, {\sl Bosonization
and Strongly Correlated Systems} (Cambridge University Press,
Cambridge, 1998); T. Giamarchi, {\it Quantum Physics in One
Dimension} (Oxford University Press, Oxford, 2004).


\bibitem{keldysh}
L.~V. Keldysh, Zh. Eksp. Teor. Fiz. {\bf 47},  1515  (1964) [Sov.
Phys. JETP {\bf 20}, 1018 (1965)]; H. Kleinert, {\it Path
Integrals in Quantum Mechanics, Statistics, and Polymer Physics}
(World Scientific, Singapore, 1995).

\bibitem{chamon96} C.deC. Chamon, D.E. Freed, and X.G. Wen, Phys. Rev. B {\bf 53}, 4033 (1996).


\bibitem{KaneFisher}
C.L. Kane and M.P.A. Fisher, Phys. Rev. B {\bf 46}, 15233 (1992).


\bibitem{PRBlong} F. Dolcini, B. Trauzettel, I. Safi, and H.
Grabert, Phys. Rev. B {\bf 71}, 165309 (2005).

\bibitem{Safi-Saleur} I. Safi and H. Saleur, Phys. Rev. Lett. {\bf 93}, 126602 (2004).

\bibitem{kane94}
C.L. Kane and M.P.A. Fisher, Phys.~Rev.~Lett. {\bf 72}, 724
(1994); V.V. Ponomarenko and N. Nagaosa, Phys. Rev. B {\bf 60},
16865 (1999); B. Trauzettel, R. Egger, and H. Grabert, Phys. Rev.
Lett. {\bf 88}, 116401 (2002).

\bibitem{PRL03} F. Dolcini, H. Grabert, I. Safi, and B. Trauzettel
Phys. Rev. Lett. {\bf 91}, 266402 (2003).

\bibitem{safi95}
I. Safi and H.J. Schulz, Phys. Rev. B {\bf 52}, R17040 (1995).

\end{thebibliography}
\end{document}